\documentclass[useAMS,usenatbib]{mn2e}
\usepackage{graphicx}

\newcommand{\1}{\Omega_{\rm M0} }
\newcommand{\2}{\Omega_{\rm X0} }
\newcommand{\3}{w_{\rm X}}
\newcommand{\4}{\Omega_{\Lambda 0} }

\newcommand{\beq}{\begin{equation}}
\newcommand{\eeq}{\end{equation}}
\newcommand{\lab}{\label}

\title[Cosmological parameters from lensing and X-ray observations]{Determining
cosmological parameters from X-ray measurements of strong lensing
clusters}
\author[M. Sereno and G. Longo]{M. Sereno$^{1,2,3}$\thanks{E-mail:
sereno@na.infn.it (MS)} and G. Longo$^{1,3}$
\\
$^{1}$Dipartimento di
Scienze Fisiche, Universit\`{a} degli Studi di Napoli ``Federico II", Via
Cinthia, Monte S. Angelo, 80126 Napoli, Italia
\\
$^{2}$Istituto Nazionale di Astrofisica - Osservatorio Astronomico di
Capodimonte, Salita Moiariello, 16, 80131 Napoli, Italia
\\
$^{3}$Istituto Nazionale di Fisica Nucleare, Sez. Napoli, Via Cinthia,
Monte S. Angelo, 80126 Napoli, Italia}
\begin{document}

\date{Accepted 2004; Received 2004; in original form 2004}

\pagerange{\pageref{firstpage}--\pageref{lastpage}} \pubyear{????}

\maketitle

\label{firstpage}

\begin{abstract}
We discuss a new method which is potentially capable of constraining
cosmological parameters using observations of giant luminous arcs in
rich X-ray clusters of galaxies. The mass profile and the mass
normalization of the lenses are determined from X-ray measurements.
The method also allows to probe the amount and equation of state of
the dark energy in the universe. The analysis of a preliminary sample
of 6 luminous, relatively relaxed clusters of galaxies strongly
favours an accelerating expansion of the universe. Under the
assumption that the dark energy is in the form of a cosmological
constant, the data provide an estimate of $ \4 = 1.1$ with a
statistical error of ${\pm} 0.2$. Including the constraint of a flat
universe and an equation of state for the dark energy $\3 \geq -1$, we
obtain $\1 = 0.10 {\pm} 0.10$ and $\3 = -0.84 {\pm} 0.14$. Relaxing the prior
on $\3$, we find that the null energy condition is satisfied at the
3-$\sigma$ confidence level.
\end{abstract}


\maketitle

\section{Introduction}
A growing body of evidence suggests that most of the energy density in
the universe consists of some sort of dark energy. Recent observations
of the power spectrum of fluctuations in the cosmic microwave
background (CMB; \shortcite{spe+al03}), of large scale structure
\citep{haw+al03} as well as supernovae data \citep{kno+al03,ton+al03}
demand for significant pressure-negative dark energy that is not
clustered with the galaxies.

The phenomenological equation of state $\3 \equiv p_{\rm X}/\rho_{\rm
X}$, i.e. the ratio between the pressure $p_{\rm X}$ of the unknown
energy component and its rest energy density $\rho_{\rm X}$ has
therefore assumed a central role in observational cosmology. The
simplest explanation for dark energy is a cosmological constant, for
which $\3
= -1$; another possibility is quintessence, i.e., a scalar field rolling
down an almost flat potential \citep{cal+al98,ra+pe98,ru+sc02};
$k$-essence, namely a scalar field with non canonical kinetic terms
\citep{arm+al99}, or models based on branes and extra dimensions, such
as the Cardassian scenario \citep{zhu+al03}, can also drive an
accelerated expansion.

Even though the equation of state is generally time-dependent, a
properly calculated weighted mean of $\3$ can approximate the late
time behavior of a large class of models \citep{wan+al00}. In a
Friedmann-Lema\^{\i}tre-Robertson-Walker (FLRW) model of universe, the
distance depends on $\3$ only through a multiple integral on the
redshift \citep{mao+al01}, and current data do not enable to
reconstruct the time evolution of the equation of state. In what
follows, we will consider an equation of state for the dark energy
assumed to be constant with redshift.

Investigations on $\3$ are usually restricted to a limited range by
assuming that the so called null energy condition (NEC) of general
relativity is fulfilled on macroscopic scales. Under quite general
hypotheses, NEC puts a lower bound to the value of $\3$
\citep{sch+al03}. While no strict mathematical proofs of NEC exist,
from a phenomenological point of view this assumption is not
justified.

The necessity of extending dark energy analyses is further illustrated
by some models of quintessence with a time dependent $\3 (t) \geq -1$
in which the assumption of constant $\3$ leads to an estimated value
of the mean equation of state smaller than $-1$ \citep{mel+al03}. So,
it is useful to enlarge the parameter space to $\3 < -1$
\citep{ha+mo02,mel+al03,sch+al03,kno+al03,spe+al03,zhu+al04}.

Matter with $\3 < -1$ is named phantom energy \citep{cal+al03}.
Theoretical motivations for such an unusual behavior have already been
proposed, cf. for instance the string theory \citep{cal+al03}. The
positive phantom energy density becomes infinite in a finite time,
overcoming all other forms of matter and leading to a big rip.

Although several independent methods of observations all converge
towards a coherent picture, the real signature of the accelerated
expansion of the universe is provided by the apparent magnitude versus
redshift test for type Ia supernovae. Other attempts to build the
Hubble diagram, like those based on the compact radio source angular
size \citep{ch+ra03}, give constraints that, even though consistent
with supernovae data, are much weaker. Systematic effects, such as
luminosity evolution, gray intergalactic dust, gravitational lensing
or selection biases, could mimic the effects of dark energy
\citep{ser+al02,ton+al03}. Although it is believed that these
contaminants are under control, an independent constraint can improve
the statistical significance of the statement about the expansion of
the universe: it is therefore still useful to develop new tools for
the determination of the cosmological parameters. Gravitational
lensing studies can provide such methods.

Results from the statistical strong gravitational lensing of
flat-spectrum radio sources based on the CLASS sample \citep{cha03}
are in good agreement with those from type Ia supernovae. Here, we are
interested in cluster of galaxies acting as lenses. Clusters of
galaxies provide a laboratory for studying and measuring the energy
content of the universe in a variety of ways. They are the largest
virialized structures and their importance in observational cosmology
is well known. In the beginning of the last century, the need for
unseen dark matter was first stated by direct estimates of their total
masses \citep{zwi33}. After a century, astronomers face again
darkness. The turn of dark energy has came.

Clusters of galaxies acting as lenses on background high redshift
galaxies have been long proposed as sources of information on the
universe
\citep{pa+go81,br+sa92,for+al97,li+pi98,lo+be99,gau+al00,men+al04}.
Values of a particular ratio of angular diameter distances can be
determined once the modeling of the lens is constrained and both the
arc position and its redshift are measured \citep{ch+ta02,gol+al02}.
\citet{io02cl} discussed the feasibility of using these gravitational
lens systems to discriminate between accelerating or decelerating
models of universe and to probe the amount and equation of state of
the dark energy. The observations of a suitable number of lensing
clusters at intermediate redshift can determine the equation of state.

The combined analysis of independent observations of clusters of
galaxies at different wavelengths can greatly improve the knowledge of
the system allowing to gain an insight into the second-order
cosmological parameters \citep{def+al04}. \citet{io03sz} used a
combined analysis of the intracluster medium (ICM) within the
hydrostatic region, as derived from both Sunyaev-Zel'dovich effect
measurements and X-ray images, to constrain the Hubble constant  and
the cosmological matter density.

Under suitable hypotheses, in a cluster of galaxy, X-ray and lensing
observations probe the same potential well. Here, we propose the use
of X-ray spectroscopic and surface brightness analyses of clusters of
galaxies to determine the mass profile and the mass normalization. In
Sec.~\ref{isot}, we show how the mass profile of a cluster can be
derived from X-ray measurements. Section~\ref{crit} discusses some
threshold requirements for strong lensing events in X-ray clusters. In
Sec.~\ref{samp}, we present the cluster data sample used in the
statistical analysis and the selection criteria applied.
Section~\ref{data} is devoted to deriving cosmological parameters and
testing the NEC condition. A discussion on some systematics that can
affect the method is contained in Sec.~\ref{syst}. Section~\ref{disc}
contains some final considerations.

\section{Isothermal $\beta$-model}
\label{isot}

At least in theory and under some simple assumptions, the distribution
of the cluster total mass can be inferred from the modeled gas
pressure distribution. In fact, the ICM is usually assumed to be
isothermal and in hydrostatic equilibrium in this potential, whereas
nonthermal processes do not contribute significantly to the gas
pressure. Under the assumption of spherical symmetry, we have
\beq
\lab{beta1}
- \frac{G M(r)}{r^2}=\frac{k_{\rm B} T_{\rm X}}{\mu m_{\rm p}} \frac{d}{dr} \ln n_{\rm e}(r),
\eeq
where $M(r)$ is the total mass of a cluster within radius $r$; $T_{\rm
X}$ and $n_{\rm e}(r)$ are the gas temperature and number density,
respectively; $m_{\rm p}$ is the proton mass; $\mu$ denotes the mean
molecular weight, assumed to be constant throughout the gas, so that
the electron number density traces the gas density. Its value is
usually fixed with solar metallicity measurements and we shall
therefore assume $\mu=0.585$, with an error of $4$\%.

A conventional $\beta$-model \citep{cav+fus76,cav+fus78}, which is
widely adopted to fit the density profile of ICM, is:
\beq
\lab{beta2}
n_{\rm e}(r) = n_{\rm e0}\left( 1+ \frac{r^2}{r_{\rm c}^2}
\right)^{- 3 \beta_{\rm X}/2},
\eeq
where $n_{\rm e0}$ is the central electron number density, $r_{\rm c}$
the core radius and the parameter $\beta_{\rm X}$ determines the
slope. By assuming a constant X-ray spectral emissivity,
Eq.~(\ref{beta2}) predicts the following X-ray surface brightness,
\beq
\label{beta2bis}
b_{\rm X} (\theta) = b_{\rm X0 } \left( 1+ \frac{\theta^2}{\theta_{\rm
c}^2}
\right)^{-3 \beta_{\rm X} +1/2},
\eeq
which provides a good fit to X-ray observations; $\theta\left(=r/
D_{\rm d}\right)$, where $D_{\rm d}$ is the angular diameter distance
to the cluster, is a dimensionless angular variable.

Inserting Eq.~(\ref{beta2}) in Eq.~(\ref{beta1}), we get the total
mass of the isothermal $\beta$-model,
\beq
\label{beta3}
M(r)= \frac{3 k_{\rm B} T_{\rm X} }{G \mu m_{\rm p}} \frac{r^3}{r_{\rm
c}^2 + r^2}.
\eeq
The mass density distribution is given by,
\beq
\label{beta4}
\rho (r) = \frac{1}{4 \pi} \frac{1}{r^2} \frac{d}{dr}M(r) = \rho_0 \frac{ 3+ \left( \frac{r}{r_{\rm c}} \right)^2}{\left[ 1 + \left( \frac{r}{r_{\rm c}} \right)^2\right]^2},
\eeq
where
\beq
\label{beta5}
\rho_0 = \frac{3 k_{\rm B} }{4 \pi G \mu m_{\rm p} }
\frac{ T_{\rm X} \beta_{\rm X} }{ r_{\rm c}^2 }.
\eeq
The surface mass density, projected along the line of sight, is
\beq
\label{beta6}
\Sigma (\theta) = \Sigma_0 \frac{ 1+ \frac{1}{2} \left( \frac{\theta}{\theta_{\rm c}} \right)^2}{ \left[ 1+ \left( \frac{\theta}{\theta_{\rm c}} \right)^2\right]^{3/2}},
\eeq
where $\Sigma_0$ is the central surface density,
\beq
\label{beta6bis}
\Sigma_0 = \frac{3}{2} \frac{k_{\rm B} }{G \mu m_{\rm p} }
\frac{T_{\rm X} \beta_{\rm X}}{\theta_{\rm c} }
\frac{1}{ D_{\rm d} }.
\eeq
The mass enclosed by a cylinder of radius $\theta$ is,
\beq
M_{\rm cyl}(\theta)= 2 \pi \Sigma_0 \left( \theta_{\rm c}D_{\rm d}
\right)^2 \frac{ \left( \frac{\theta}{\theta_{\rm c}} \right)^2}{
\left[ 1+ \left( \frac{\theta}{\theta_{\rm c}}
\right)^2\right]^{1/2}},\eeq

The profile in Eq.~(\ref{beta6}) is known as softened power-law
surface mass density and the slope parameter is fixed to $1/2$
\citep{sef}.

In a spherically symmetric, regular lens, a strong lensing event can
occur if \citep{sef}:
\beq
\label{beta6ter}
\Sigma_0 > \Sigma_{\rm cr},
\eeq
where $\Sigma_{\rm cr}$ is the critical surface mass density,
\beq
\lab{eq5}
\Sigma_{\rm cr} \equiv \frac{c^2}{4 \pi G} \frac{D_{\rm s}}{D_{\rm d} D_{\rm ds}},
\eeq
being $D_{\rm ds}$ the angular diameter distance from the lens to the
source and $D_{\rm s}$ the angular diameter distance from the observer
to the source. A tangential critical curve appears at $\theta_{\rm t}$
\citep{sef},
\beq
\lab{beta7}
\theta_{\rm t} = \theta_{\rm c} \left[ \left( \frac{\Sigma_0}{\Sigma_{\rm cr}} \right)^2 -1\right]^{1/2};
\eeq
a radial critical curve forms at
\beq
\theta_{\rm r} = \theta_{\rm c} \left[ \left( \frac{\Sigma_0}{\Sigma_{\rm cr}} \right)^{2/3} -1 \right]^{1/2},
\eeq
The corresponding radial caustic in the source plane has an angular
radius $\eta_{\rm r}$, where
\beq
\eta_{\rm r} = \frac{\theta_{\rm r}^3}{\theta_{\rm c}^2}.
\eeq
A source inside the radial caustic produces three images, one which
instead is outside has only one image.

The mass profile of a cluster can be fully determined by X-ray
measurements. A morphological analysis of the X-ray surface brightness
allows to constrain the parameters $\theta_{\rm c}$ and $\beta_{\rm
X}$, while the gas temperature $T_{\rm X}$ is determined with
spectroscopic measurements.

\section{Strong lensing criteria}
\label{crit}

\begin{figure}
        \resizebox{\hsize}{!}{\includegraphics{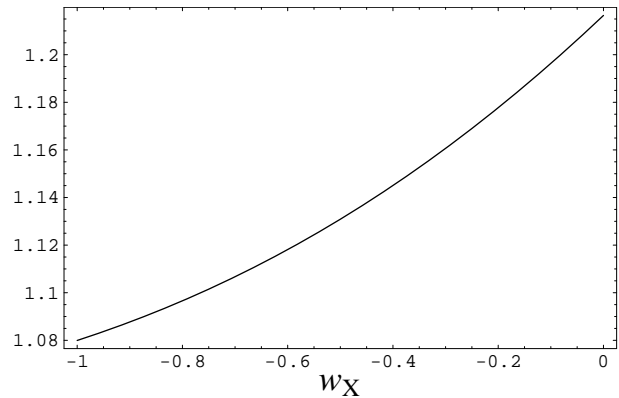}}
        \caption{The ratio of distances $D_{\rm s}/D_{\rm ds}$ as a function of
        $\3$ in a flat model of Universe with $\1 = 0.2$. Here, we assume
        $z_{\rm d}=0.3$ and $z_{\rm s}=1$. }
        \label{ratio_wx}
\end{figure}

\begin{figure}
        \resizebox{\hsize}{!}{\includegraphics{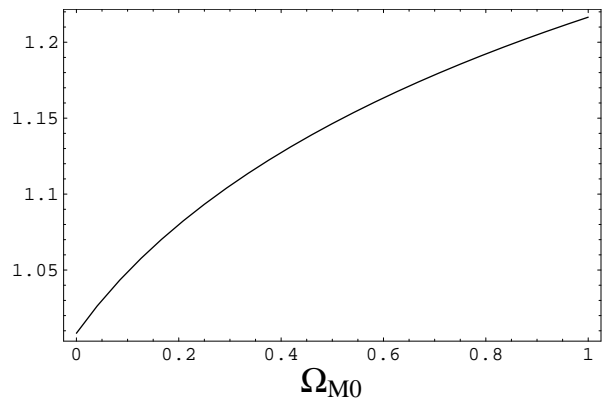}}
        \caption{The ratio of distances $D_{\rm s}/D_{\rm ds}$ as a function of $\1$
        in a flat model of Universe with cosmological constant ($w_{\rm X}=-1$). Here,
        we assume $z_{\rm d}=0.3$ and $z_{\rm s}=1$.}
        \label{ratio_om}
\end{figure}

As seen in the previous section, when we assume a spherical mass
distribution, a lens can produce multiple image systems only if the
central surface density overcomes the critical one. According to
Eqs.~(\ref{beta6bis}~,\ref{eq5}), the criterium in
Eq.~(\ref{beta6ter}) is fulfilled if
\beq
\label{crit2}
\left( \frac{k_{\rm B} T}{11~{\rm KeV}} \right) \left( \frac{\beta_{\rm X}}{0.6} \right)
\left( \frac{30\arcsec}{ \theta_{\rm c}} \right) \stackrel{>}{\sim} \frac{D_{\rm s}}{D_{\rm d
s}}.
\eeq
In other words, as it was to be expected, very hot and massive
clusters with a small core radius are more likely to form arc-like
systems.

A proper X-ray modelling is needed to check the condition in
Eq.~(\ref{crit2}). Estimating the temperature profile requires a
detailed spatially resolved spectroscopy. A multiphase structure in
the central ICM could result in significant discrepancies between
emission-weighted and mass-weighted estimates of the temperature. A
simple isothermal model which does not incorporate the cooling-flow
components or other complexities in the X-ray structure can lead to
systematic underestimates of order 10-40\% in the temperature and mass
determinations \citep{all98}.

On the average, detailed X-ray morphologies show rapid variations in
the central regions; such variations may be due, for instance, to
recent merger or accretion activities and, therefore, they might
invalidate the assumption of hydrostatic equilibrium. Complex
dynamical activities can lead to overestimating the core radius of the
dominant clump in the cluster, as inferred from X-ray data. The core
radius for the gravitating matter in cooling-flow clusters is usually
small, in agreement with optical studies. The radius nearly doubles
for intermediate systems, and, in the case of non-cooling-flows, it
may be larger by a factor six \citep{all98}.

Threshold values for strong lensing events depend on the cosmological
parameters through the ratio of distances $D_{\rm s}/D_{\rm ds}$. We
consider a FLRW universe filled with non-interacting pressureless
matter and dark energy, described as $X$-matter
\citep{chi+al97,tu+wh97}. Since the contribution from relativistic
particles is negligible in the redshift range investigated in our
analysis, we will neglect it in what follows. In such a model of
universe, the angular diameter distance between an observer at $z_{\rm
d}$ and a source at $z_{\rm s}$ is
\beq
\label{crit3}
D(z_{\rm d}, z_{\rm s})=\frac{c}{H_0}\frac{1}{1+z_{\rm
s}}\frac{1}{|\Omega_{\rm K0}|} {\rm Sinn} \left( \int_{z_{\rm
d}}^{z_{\rm s}} \frac{|\Omega_{\rm K0}|}{{\cal E}(z)} dz \right),
\eeq
where
\begin{eqnarray}
{\cal E}(z) & \equiv & \frac{H(z)}{H_0} \\ & =& \sqrt{ \1 (1+z)^3+\2
(1+z)^{3(w_{\rm X}+1)} +\Omega_{\rm K0}(1+z)^2 },\nonumber
\end{eqnarray}
and $H_0$ is the present value of the  Hubble parameter; $\1$ and $\2$
are the today normalized densities of dust and dark energy,
respectively; $\Omega_{\rm K0} \equiv 1- \1 -\2$; Sinn is defined as
being sinh when $\Omega_{\rm K0}>0$, sin when $\Omega_{\rm K0}<0$, and
as the identity when $\Omega_{\rm K0}=0$. For the expression of the
distance in an inhomogeneous universe, we refer to
\citet{ser+al01,ser+al02}.

Depending on the values of the cosmological parameters, significant
variations in the distance ratio have to be expected. As it can be
seen from Figs.~(\ref{ratio_wx},~\ref{ratio_om}), the ratio $D_{\rm
s}/D_{\rm ds}$ is an increasing function of both $\1$ and $\3$. Dark
energy with large negative pressure helps the formation of arc-like
systems in X-ray clusters. The best case is that of a cosmological
constant. Changing $\3$ from $-1$ to $0$, the ratio of distances
increases of 13\% for a lens at $z_{\rm d}=0.3$ and a background
source at $z_{\rm s}=1$ in a flat universe with $\1 =0.2$.
Sub-critical values of the pressureless matter density also favour
strong lensing systems. The relative variation from $\1 =0$ to $\1 =1$
is $\sim 21\%$ for $\3
=-1$. A large value of a cosmological constant increases the cross
section for strong lensing and lowers the threshold value of the
combination of X-ray parameters in Eq.~(\ref{crit2}).

\section{Cluster sample}
\label{samp}

\begin{table*}
\caption{\label{tab_clus} Information on arcs and X-ray properties of the strong
lensing cluster sample.  $^a$ References: (1) \citet{all+al01};  (2)
\citet{sch+al04}; (3) \citet{don+al03}; (4) \citet{ara+al02};  (5)
\citet{all+al01b}; (6) \citet{mol+al99}; (7) \citet{hic+al02}; (8)
\citet{all+al02}.}
        \begin{center}
        \begin{tabular}{|l|l|l|l|l|l|l|l|}
        \hline
        Cluster     & $z_{\rm d}$ & $z_{\rm arc}$ &  $\theta_{\rm arc}(\arcsec)$ & $T$(KeV)  & $\beta$
        & $\theta_{\rm c}(\arcsec)$  & ref$^a$ \\
        \hline
A 2390          &  0.228   & 0.913   & 38.0  & $11.5^{+1.6}_{-1.3}$ &
$0.48 {\pm} 0.02$ & $11.8 {\pm} 0.3$ & 1,2
\\
MS 0451.6-0305  &  0.539   & 2.911   & 25.9  & $10.3^{+1.2}_{-1.0}$ &
$0.70  {\pm}  0.02$    & $31  {\pm} 1.0$ & 3
\\
MS 1358.4+6245   & 0.328   & 4.92   &  21.0  & $7.16 {\pm}  0.10$       &
$0.69  {\pm}  0.02$ & $ 20.5   {\pm}  5.2$ & 4
\\
MS 2137.3-2353   & 0.313   & 1.501  &  15.3  & $5.56^{+0.46}_{-0.39}$
&  $0.63^{+0.04}_{-0.03}$ & $8.3^{+1.4}_{-1.2}$ & 5,6
\\
PKS 0745-191     & 0.103   & 0.433  & 18.2   & $9.55^{+1.06}_{0.75}$ &
$0.47 {\pm} 0.02$    & $18.0^{+0.4}_{-0.6}$ & 5,7
\\
RX J1347.5-1145  &  0.451  &  0.806 &  34.9  & $12.2 {\pm} 0.6$ & $0.57 {\pm}
0.02$  & $4.8 {\pm} 0.3$ & 8
\\
        \end{tabular}
        \end{center}
\end{table*}

Our method to determine cosmological parameters relies on a number of
assumptions: hydrostatic equilibrium, isothermality, spherical
symmetry. The selection of the sample must therefore be very careful
in order not to violate them \citep{smi+al03}. We have collected data
from literature on luminous, relatively relaxed clusters with giant
luminous arcs and well measured, X-ray spatially-resolved
observations. Then, we have selected a sub-sample suitable for our
analysis, by requiring: $i)$ good agreement of optical and/or lensing
mass measurements with the X-ray results; $ii)$ a regular X-ray
morphology; $iii)$  temperature measured with {\it Chandra}
Observatory or {\it XMM-Newton} satellites;  $iv)$ arcs with known
spectroscopic redshift.

The agreement between mass estimates from independent methods firmly
limits the systematic uncertainties when determining the cosmological
parameters. Apparently relaxed clusters can show significant
discrepancy between X-ray and lensing masses
\citep{kne+al03,ota+al04}. Together with the regularity of the X-ray
surface brightness emission, the consistency among the X-ray, optical
and/or gravitational (both weak and strong) lensing mass measurements
is a reliable indication of hydrostatic equilibrium and strongly
suggests that the global properties of the clusters can be used as
cosmological probes. X-ray gas should be supported by thermal pressure
which dominates over non-thermal processes and the assumption of
hydrostatic equilibrium can be considered valid.

Gas temperatures are typically very hard to measure \citep{gio+al04}.
whereas low resolution measurements suffer from point-source emission
contaminating cluster spectra, {\it Chandra}'s or {\it XMM-Newton}'s
spatial resolutions allow the identification of point sources and an
independent spectral determination of the temperature in each radial
shell \citep{all+al01,don+al03}. In order to be conservative in our
selection procedure, we consider only clusters for which the
temperatures determined with last generation satellites agree with
previously reported values from ROSAT and/or ASCA data. Such an
agreement is a strong indication that the temperature estimate is
substantially free from systematics and that the adopted uncertainty
is realistic.

We wish to stress that conditions $i)$ and $ii)$ are often fulfilled
by cooling flow clusters, which turn out to be quite regular,
dynamically relaxed systems. The X-ray emission is typically symmetric
and little or no substructures appear at optical wavelengths. These
clusters are also amongst the most X-ray luminous.

Furthermore, the comparison of masses derived from X-ray and strong
lensing show excellent agreement for cooling-flow clusters
\citep{all98,wu00}, while discrepancies appear in dynamically active
clusters with either small or no cooling-flows. The oversimplification
of a spherical model and the application of isothermality and
equilibrium hypotheses may be inappropriate in presence of local
dynamical activities.

Due to their regularity, from the point of view of gravitational
lensing, cooling-flow clusters are ``critical", i.e. they may form
giant arcs, only if the condition in Eq.~(\ref{crit2}) is fulfilled.
On the other hand, substructures and non-equilibrium states in
dynamically active, non-cooling flow systems enhance the probability
of detecting strong lensing events \citep{ba+st96}, which are
favourite in the presence of irregular and complex dark matter
distributions. A simple spherical model, which does not account for
the local dynamical activities, may lead to overestimating the
gravitating masses inferred from lensing studies.

The nearly coincidence of a X-ray single peak with the cD galaxy
position, which has been argued to be a good indicator of the presence
of local violent activities of the intracluster gas in the central
core, has also been proposed as an indicator of hydrostatic
equilibrium \citep{all98}. Typically, in cooling-flow clusters the
emission is sharply peaked to a position nearly coincident or very
close to the optically dominant cluster galaxy, which generally is the
centre of the matter distribution as inferred from lensing studies.

After the application of our selection criteria, our final sample
reduced to 6 clusters where strong lensing analysis and X-ray studies
are likely to trace the same gravitational potential well. The main
properties of this sample are listed in Table~\ref{tab_clus}. Our
sample, spanning the range $0.1 < z_{\rm d} <0.54$, constitutes an
updated sub-sample of the list in \citet{wu00}. Information on
positions and redshifts of arcs were taken from \citet{wu00} and
references therein. New spectroscopic redshift measurements for
MS~0451.6-0305 and MS~2137.3-2353 were instead taken from
\citet{bor+al04} and \citet{san+al02}, respectively.

For each cluster, the ICM has been modeled with an isothermal
$\beta$-model. We consider fits to the X-ray surface brightness which
extend well beyond the core radius. When more morphological analyses
are available for a single cluster, we adopt results from fits that
extend to the outer radius. Since we collected data from independent
groups, in order to avoid unrealistic small uncertainties when
comparing the data, we assume an error of $0.02$ on the slope $\beta$
when determined from {\it Chandra} observations.

Details on the X-ray spectroscopic and imaging analysis for each
cluster can be found in the references listed in Table~\ref{tab_clus}.
Clusters in our sample have been the subjects of X-ray, optical and,
often, Sunyaev-Zel'dovich effect studies. They present very regular
X-ray isophotal patterns. Despite some local substructures which have
been revealed in highly detailed X-ray surface brightness maps,
morphological analyses support the spherical approximation: there are
no evidences for violent merger shocks and the luminosity and
temperatures do not seem to have been boosted substantially
\citep{all+al01,all+al02,ara+al02,don+al03,hic+al02,mol+al99}.

Apart from MS~0451.6-0305, all the clusters in our final sample
present massive cooling flows in the central regions. Furthermore, for
each cluster except again MS~0451.6-0305, X-ray morphological analyses
have revealed a sharp central surface brightness peak at a position
nearly coincident, within the astrometric resolution of {\it Chandra}
observations, with the optical centroid for the dominant cluster
galaxy. In MS~0451.6-0305, whereas the soft-energy peak of the X-ray
surface brightness lies on the brightest cluster galaxy, a harder peak
is slightly shifted \citep{don+al03}. However, due to the consistency
between the X-ray and Sunyaev-Zel'dovich estimates of the central
electron density and the agreement between the X-ray, lensing and
optical estimates of the total mass of this cluster \citep{don+al03},
we have included it in our final sample.

\section{Data analysis}
\label{data}

\begin{figure}
        \resizebox{\hsize}{!}{\includegraphics{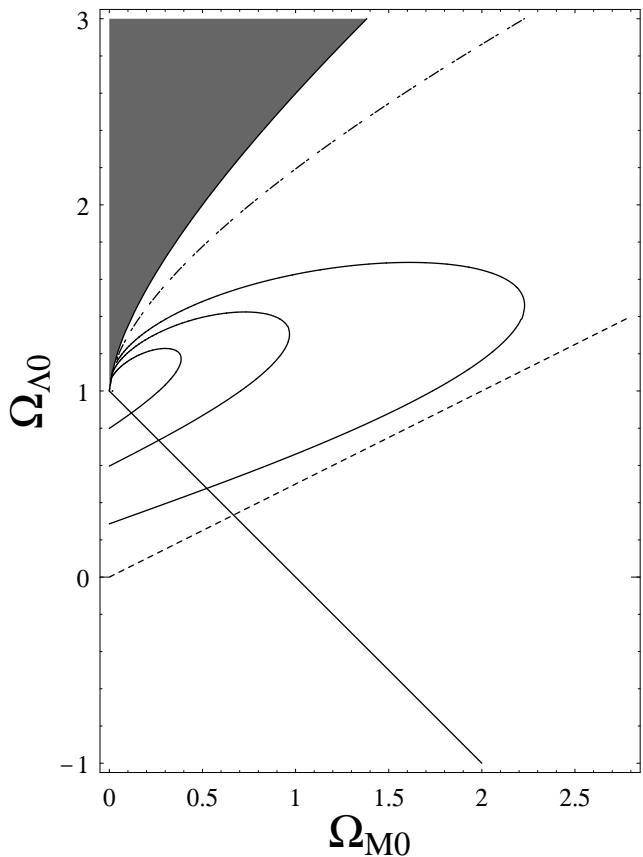}}
        \caption{Contours of constant $\Delta \chi^2$ in the $\1$-$\4$ plane.
        Lines are traced for $\Delta \chi^2$ values of 2.30, 6.17 and 11.8. Accelerated
        models of universe
        are above the dashed line. The full line represents the locus of flat models
        of universe ($\Omega_{\rm K}=0$); the long-dashed line represents models with a
        null distance to $z_{\rm s}=4.92$; bouncing models in the upper-left shaded region
        do not have big bang.}
        \label{om_ol_plane}
\end{figure}

\begin{figure}
        \resizebox{\hsize}{!}{\includegraphics{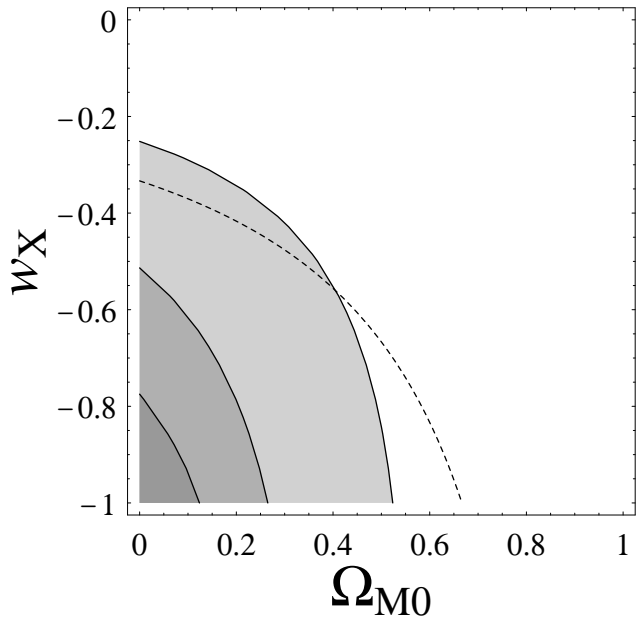}}
        \caption{Confidence contours on the $\1$-$\3$ plane. The shaded regions show
        the joint 1, 2 and 3-$\sigma$ confidence regions, bounded by $\Delta \chi^2$ values
        of 2.30 and 6.17 and 11.8, respectively. Accelerated models of universe are below
        the dashed line.}
        \label{om_wx_plane}
\end{figure}

Let us now perform a $\chi^2$ statistics. Observations of giant arcs
in X-ray clusters enable us to estimate the distance ratio $D_{\rm
ds}/D_{\rm s}$. From Eqs.~(\ref{beta6bis},~\ref{eq5},~\ref{beta7}), we
get
\beq
\label{chi1}
\left .\frac{D_{\rm ds}}{D_{\rm s}}\right|_{\rm obs} =
\frac{\mu m_{\rm p}c^2}{6 \pi}\frac{1}{\beta_{\rm X}
T_{\rm X}}\sqrt{\theta^2_{\rm t}+\theta^2_{\rm c}}.
\eeq
The parameters $T_{\rm X}$, $\beta_{\rm X}$ and $\theta_{\rm c}$ are
derived from a X-ray data analysis of temperature and surface
brightness profiles.

The value of the angular radius of the tangential critical curve was
derived from the observed arc radius $\theta_{\rm arc}$. According to
\citet{ono+al99}, a conventional analysis where $\theta_{\rm
t}=\theta_{\rm arc}$, could overestimate the strong lensing mass by
10-30\%. To correct for this effect, we take $\theta_{\rm t}=\epsilon
\theta_{\rm arc}$, with $\epsilon = \left( 1/\sqrt{1.2}
\right) \pm 0.04$.

The $\chi^2$ reads
\beq
\label{chi2}
\chi^2 = \sum_{\rm  systems} \left\{ \frac{
\left .\frac{D_{\rm ds}}{D_{\rm s}}\right|_{\rm obs}^i -
\frac{D_{\rm ds}}{D_{\rm s}}(z_{\rm d}^i,z_{\rm s}^i;\1,\2,\3) }{\sigma_i}
\right\}^2
\eeq
The statistical errors $\sigma_i$ include uncertainties arising from
$T_{\rm X}$, $\beta_{\rm X}$, $\theta_{\rm c}$, $\mu$ and $\theta_{\rm
arc}$. Many input data in our analysis are presented with asymmetric
uncertainties. To obtain unbiased estimates, we apply correction
formulae for mean and standard deviation as given by Eqs.~(15,~16) in
\cite{dag04}. Since the analysis of the X-ray parameters is highly
correlated, we add the uncertainties from $\beta_{\rm X}$ and
$\theta_{\rm c}$ as maximum errors.

Once we assume that the $\chi^2$ description of the data is a good
approximation, the likelihood of the cosmological parameters turns out
to be ${\cal L} (\1, \2, \3) \propto e^{-\chi^2(\1,
\2, \3)/2}$. Following Bayesian statistics, the posterior
probability of a particular model after analyzing the available data
is proportional to the prior probability of that model multiplied by
the likelihood. For each parameter in the model, its one dimensional
posterior probability function is found by marginalizing over all
other parameters. Unless otherwise noted, for each parameter we quote
the one-dimensional mean and standard deviation.

We first consider the case of a cosmological constant $(\2 =\4,
\3 =-1$). We limit our analysis to the region in the parameter space
for which the distance to the most distant arc in our sample (at
$z_{\rm s}=4.92$) is not null. In Fig.~\ref{om_ol_plane}, we show the
$\chi^2$ function. According to the maximum likelihood ratio method,
joint 1, 2 and 3-$\sigma$ confidence regions for two parameters are
bounded by $\Delta \chi^2$ values of 2.30 and 6.17 and 11.8,
respectively. As it can be seen, our method is quite sensitive to the
cosmological constant, while the pressureless matter density is poorly
constrained. The best-fitting model, $(\1,\4) \simeq (0.01, 0.99)$, is
nearly coincident with a de Sitter model. As it was discussed in
\citet{got+al01}, appropriate priors should be as vague or
noninformative as possible. We consider two types of priors: $i)$ a
uniform prior, with the prior probability proportional to $ d\1 d\4$;
$ii)$ the logarithmic or Jeffrey prior, appropriate for a positive
variable like $\1$, with a prior probability proportional to $ d
\4 d\1 /\1 $. Results on $\1$ are highly sensitive on the choice of the prior,
indicating that better data are needed to convincingly constrain this
parameter \citep{got+al01}. On the other hand, the determination of
the cosmological constant is nearly insensitive to the prior.

Let us consider an uniform prior. Within 3-$\sigma$ statistical
significance, the data strongly favour accelerating models, whereas an
Einstein-de Sitter model of universe ($\1 =1, \Omega_{\rm K0}=0$) is
not consistent. After marginalization, we find $\4 = 1.1 {\pm}0.2$.

Then, following theoretical prejudices on an inflationary, nearly flat
model of universe with a positive cosmological constant, we restrict
our analysis to the square region of the parameter space defined by
$0\leq \1 \leq 1$ and $0 \leq \4 \leq 1$.  Adopting a uniform prior,
we find $\1 = 0.2 {\pm}0.2$ and $\4 =0.87 {\pm} 0.12$. Whichever prior we use,
our analysis provides a strong evidence supporting the existence of
dark energy.

Let us now analyse the dark energy equation of state. We consider flat
models ($\Omega_{\rm K}=0$). As a first step, we impose a prior on the
equation of state, $-1 \leq \3 \leq 0$. The best fit model occur for
$(\1, \3) \simeq (0.01, -1)$. Adopting uniform priors, after
marginalization, we get $\1 = 0.10 {\pm} 0.10$ and $\3 = -0.84 {\pm}0.14$;
accelerating models of universe are favoured at the 2-$\sigma$ level,
see Fig.~\ref{om_wx_plane}. Our analysis favours a model of universe
in accelerate expansion with the energy budget dominated by dark
energy. At this point, with the prior on $\3$, the cosmological
constant turns out to be the best candidate as dark energy.

Our results are in a substantial agreement with current determinations
\citep{ha+mo02,mel+al03,kno+al03,spe+al03,sch+al03,ton+al03,zhu+al04}.
A combined analysis of supernovae, galaxy distortion from the
Two-Degree Field Galaxy Redshift Survey (2dFGRS; \citet{haw+al03}) and
CMB data from WMAP \citep{spe+al03} provide  an estimate of $\1
=0.27^{+0.06}_{-0.05}$ and a 95\% upper confidence limit of $\3 <0.78$
\citep{kno+al03}. An independent analysis in \citet{spe+al03}, with
some different external constraints and combining CMB, 2dFGRS power
spectrum, supernovae, and the HST Key Project data, also gives the
same constraint on $\3$.

\subsection{Energy conditions violations}

\begin{figure}
        \resizebox{\hsize}{!}{\includegraphics{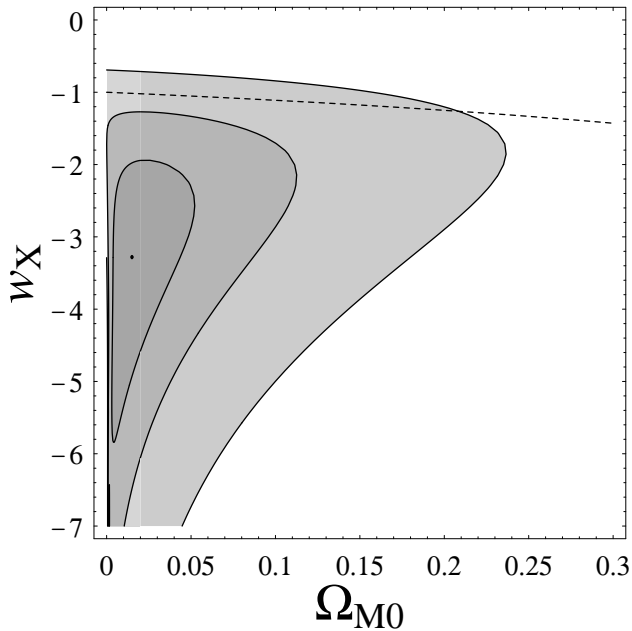}}
        \caption{Confidence contours on the $\1$-$\3$ plane. Shaded regions are as in
        Fig. \ref{om_wx_plane}. The dot indicates the best fit model. The NEC
        condition is fulfilled above the dashed line.}
        \label{om_wx_plane_NEC}
\end{figure}

Under quite general hypotheses, in the case of a spatially flat FLRW
geometry, with a negligible contribution from relativistic particles,
NEC puts a lower bound on $\3$, $\3 \geq -(1-\1)^{-1}$
\citep{sch+al03}. For $\1 =0$, we have $\3 \geq -1$, i.e., the usual
limit assumed in most of the investigations.

Following \citet{sch+al03}, we use $\3$ to test NEC. We find a best
fitting value of $\3 \simeq -3.28$. Whereas at 2-$\sigma$ confidence
level the NEC is violated, at the 3-$\sigma$ confidence limit, it is
fulfilled, see Fig.~\ref{om_wx_plane_NEC}. The case of a cosmological
constant is consistent with data at the 3-$\sigma$ level.

Distance data alone do not well bound from below the equation of state
and the lower limit of $\3$ is quite uncertain \citep{kno+al03}. The
weight of probability at very low $\3$ pulls the confidence level
downward. From supernovae data, at 99\% confidence level, an upper
limit of $\3 <-0.64$ is derived, but interval extends to $\3 < -10$
\citep{kno+al03}. The degeneracy can be broken by using orthogonal
constraints, such as those arising from abundances of nearby clusters
of X-ray clusters \citep{sch+al03}. The analysis in \citet{kno+al03}
provides a measurement of $\3 =-1.05^{+0.15}_{-0.20}(\rm
statistical)\pm 0.09$ (identified systematics). The independent
analysis in \citet{spe+al03} gives a similar result, $\3 =-0.98 \pm
0.12$.

\section{Systematics}
\label{syst}

In the previous section, we have performed a statistical analysis
based on data found in the literature. Whereas we selected a regular
sample, with well measured properties, a number of systematics can
still plague our analysis. We want now to address how they can affect
our results.

\subsection{Density profile}

We have our high-redshift sample by using a $\beta$-profile.
Application of such a model provides robust estimates of the total
mass, as shown for simulated clusters \citep{moh+al99} and observed
clusters \citep{ett+al04}. It is still interesting, however, to
investigate the impact of a central cusp in the dark matter
distribution, as predicted by numerical simulations, on our estimates.
To this aim, we consider the so called universal Navarro, Frenk and
White (NFW) density profile \citep{nav+al95}. Since the similarity
between the distribution of ICM tracing the dark halo of the NFW
potential and the $\beta$-profile \citep{mak+al98,et+fa99}, NFW
parameters can be derived from the $\beta$-model ones with helpful
empirical formulae. We therefore applied the relations found in
\cite{mak+al98}. The introduction of a central cusp slightly reduces
the estimated value of the cosmological constant of $\sim 0.07$ and
the best fit value of $\3$ increases of $\sim 0.3$.

\subsection{Temperature}

Isothermal models provide a good representation of the physical
properties of galaxy clusters \citep{ett+al02}. Whereas the presence
of a gradient in the temperature profile would reduce the total mass
measurements and increase, consequently, the derived value of the
cosmological constant, expected gradients in clusters are not steep
enough to affect significantly the estimated total mass
\citep{ett+al02}. However, in order to be more confident on the
temperature values used in our analysis, we checked our sample for
consistency of the new temperature measurements obtained with high
resolution satellite with those previously reported from ROSAT and/or
ASCA data, covering a larger field of view (see data in Table~2 in
\cite{wu00}). This should control and exclude any systematic bias. In
any case, a systematic offset in the temperature estimate of $\sim
5\%$ would determine a variation in $\Omega_{\Lambda 0}$ of $\sim
0.15$, with a positive offset determining an overestimation of the
cosmological constant. The constraint on $\3$, when no priors are
adopted, would be nearly washed out by such an offset.

\subsection{Ellipticity and substructures}

One of the more important sources of indetermination comes from the
modeling of the mass profile of the lens. Morphological X-ray analyses
revealed a tendency for clusters of galaxies to be elliptical, with a
mean projected axis ratio of $\sim 1.25$ \citep{moh+al95,def+al04}.
Furthermore, together with the overall mass profile, sub-structures
should also be considered. High detailed X-ray surface brightness maps
can reveal interesting details in the distribution of ICM. In
RX~J1347.5-1145 \citep{all+al02}, {\it Chandra} observations revealed
evidences of shocked gas to the south-east of the main X-ray peak,
coincident with a region of enhanced Sunyaev-Zel'dovich effect,
probably resulting from recent subcluster merger activity. Presence of
some substructure in A~2390 \citep{all+al01} and PKS~0745-191
\citep{hic+al02} was also detected and a hint that the brightest
cluster galaxy may be just settling into the cluster core has been
reported for MS~0451.6-0305 \citep{don+al03}.

However, when the lens presents a rather regular morphology, even if a
``not correct" potential shape is used in the reconstruction, or the
contribution of small sub-structures is neglected, the cosmological
parameters are still retrieved, although with somewhat larger errors
\citep{gol+al02}.

To best investigate the role played by these deviations from the
assumed projected cylindrical symmetry, we considered an additional
source of systematic uncertainty in the estimation of the position of
the tangential critical radius, of the order of $\sim 20\%$. The
effect is quite large, introducing an error of $\sim 0.15$ in the
determination of the reduced cosmological constant, and of $\sim 1.5$
in the determination of the global minimum for the equation of state
$\3$ (when no prior is assumed).

\section{Discussion}
\label{disc}

We discussed a method to probe the cosmological parameters, based on
observations of giant luminous arcs in strong lensing clusters whose
mass scales and mass profiles are determined with X-ray spectroscopic
and surface brightness analyses.

Giant arcs are more likely to form in massive, hot clusters with a
small core radius. A large contribution to the energy budget of the
universe in the form of dark energy with strong negative pressure also
helps the occurrence of phenomena of strong lensing events.

In order to constrain the cosmological parameters, we considered an
updated subsample of rich, strong lensing clusters from the list in
\citet{wu00}. Only dynamically relaxed, regular systems were included.
Our statistical analysis supports a dark energy dominated model of
universe, in agreement with combined measurement from supernovae,
galaxy clustering and CMB data, and a clear signature for an
accelerated expansion of the universe is provided at the 2-$\sigma$
confidence level. Dark energy with a strong negative pressure is also
favoured. The cosmological constant provides the strongest candidate,
but a wide range of quintessence models are compatible. The data are
consistent with the NEC at 3-$\sigma$ statistical significance.

The method that we have proposed is independent of the supernovae data
and provides an alternative tool to probe the expansion of the
universe. Our analysis is based on a very limited number of hypotheses
and is not sensitive to $H_0$. Clusters of galaxies are not required
to be standard rulers or candles. On the contrary of supernovae
analyses, our analysis does not require either homogeneity in the
sample nor absence of evolution. Moreover, contamination from
neighbours is truly negligible.

Then, the redshift range covered by observations of lensed, highly
magnified galaxies extends well beyond the supernova survey limits and
can probe the changing point between the pressureless matter dominated
era and the dark energy one.

The method however needs a detailed modeling of the lens from X-ray
observations and a deep understanding of the structure of the X-ray
potential well is crucial. The identification of irregularities and
substructures in a complex morphology is very important when
predicting arc position.

Assumptions made in the methodology must be verified testing
systematics. The use of new X-ray data enables us to refine the
criteria to select clusters in hydrostatic equilibrium, such as the
nearly coincidence of a X-ray single peak with the cD galaxy position
or, mainly, the presence of a cooling-flow. Thus far, detailed X-ray
analyses, based on {\it Chandra} or {\it XMM-Newton} observations,
have been explicitly derived for several lensing clusters. Whereas for
some clusters (A1835, MS~1358, and A2390), there is a good agreement
with the lensing analyses \citep{all+al01,sch+al01,ara+al02}, a
significant discrepancy is found, for example, in the irregular
clusters A2218 \citep{mac+al02} and A1689 \citep{xu+wu02}. Whereas
A2218 shows an offset between the X-ray centroid and the central
dominant cD galaxy in A1689 the X-ray centre perfectly coincides with
the cD galaxy. An analysis of AC~114 \citep{def+al03} shows a good
agreement between X-ray and lensing results, despite an offset of
$\sim 10\arcsec$. In a forthcoming paper, we want to extend our method
to clusters of galaxies with an irregular X-ray morphology but very
rich in multiple image systems.

\section*{Acknowledgments}
The authors thank E. De Filippis for useful discussions and the
anonymous referee for her/his comments that helped to greatly improve
and clarify several points of the paper.

\bsp

\label{lastpage}

\end{document}